\documentclass[11pt]{article}
\usepackage{epsfig}
\usepackage{amsmath,amssymb,cite}

\numberwithin{equation}{section}

\begin{document}
\begin{flushright}
UCD-2001-06\\
\end{flushright}
\begin{center}
{\large{\bf SCALAR MULTI-SOLITONS ON THE FUZZY SPHERE}}
\bigskip 

Sachindeo Vaidya\footnote{vaidya@dirac.ucdavis.edu}\\ 
{\it Department of Physics, \\
University of California, Davis, CA 95616,USA.} \\

\end{center}

\begin{abstract}
We study solitons in scalar theories with polynomial interactions on
the fuzzy sphere. Such solitons are described by projection operators
of rank $k$, and hence the moduli space for the solitons is the
Grassmannian $Gr(k,2j+1)$. The gradient term of the action provides a
non-trivial potential on $Gr(k,2j+1)$, thus reducing the moduli
space. We construct configurations corresponding to well-separated
solitons, and show that although the solitons attract each other, the
attraction vanishes in the limit of large $j$. In this limit, it is
argued that the moduli space is ${({\mathbb C}P^1)}^{\otimes k} /{\cal
S}_k \simeq {\mathbb C}P^k$. For the $k$-soliton bound state, the
moduli space is simply ${\mathbb C}P^1$, all other moduli being
lifted. We find that the moduli space of multi-solitons is smooth and
that there are no singularities as several solitons coalesce. When the
fuzzy $S^2$ is flattened to a noncommutative plane, we find agreement
with the known results, modulo some operator-ordering
ambiguities. This suggests that the fuzzy sphere is a natural way to
regulate the noncommutative plane both in the ultraviolet and
infrared.
\end{abstract}

\section{Introduction}

Theories on noncommutative spaces have been studied vigorously for a
while now. They arise in string theory in a certain corner of moduli
space \cite{seiwit}. Noncommutative theories can also be studied
independently and in their own right, and provide a variety of
interesting new phenomena. Theories on noncommutative compact
manifolds often come with an ultraviolet cut-off, and are thus
potentially regulated in their short distance behavior. However, the
quantum properties of noncommutative theories are considerably subtle:
even in a simple scalar theory on noncommutative $R^4$ there is mixing
between ultraviolet and infrared degrees of freedom, as was first
shown in \cite{mirase}.

Interestingly, noncommutative scalar theories in $(2n+1)$-dimensions
also have stable finite energy classical configurations. This first
shown in \cite{gomist}, who constructed these finite energy
configurations from projector operators and interpreted them as
solitons. In a subsequent paper \cite{gohesp}, the multi-soliton
moduli space was studied in detail.

Using the techniques first discussed in \cite{gohesp} (see also
\cite{sprvol}), we will study solitons in scalar theories defined on
$S_F^2 \times R$, where $S_F^2$ is a fuzzy sphere. A fuzzy sphere is
described by a finite dimensional algebra generated by 3 matrices
$X_a$ that satisfy
\begin{equation} 
[X_a,X_b]= i\frac{R}{\sqrt{j(j+1)}}\epsilon_{abc} X_c,  \quad X_a X_a =
R^2 {\bf 1}, \quad a,b,c=1,2,3 \quad {\rm and}\quad j \in {\mathbb Z}/2.
\end{equation} 
The $X_a$ are $(2j+1)\times(2j+1)$ matrices proportional to the
$(2j+1)$-dimensional representation of the generators $J_a$ of the
$SU(2)$ algebra. In this article, we will be interested in two limits:
in one case, we take $j\rightarrow \infty$ at fixed $R$ to get an
ordinary sphere (of radius $R$), while in the other limit we take $j,R
\rightarrow \infty$ with $R^2/j$ fixed, to get the noncommutative
plane. The multi-soliton configurations will be constructed using
$SU(2)$ coherent states, and we will look in some detail at the case
of two solitons. It will be argued that well-separated solitons are
labeled by their location on an ordinary sphere. For the class of
solitons that we consider, the solitons are found to attract each
other, as was also pointed out by \cite{sprvol}. Interestingly, the
attraction becomes very weak in the limit of large $j$ even at finite
radius, and the solitons behave as free particles. We will also argue
that the multi-soliton moduli space is smooth, and that there are no
singularities as two or more solitons coalesce.

It must be emphasized that the configurations we call solitons in
this article are different from the ones that have been studied
earlier in similar contexts \cite{gkp12,gkp3,bbvy,balvai}. The
solitons, monopoles and instantons discussed in \cite{bbvy,balvai}
were based on a discrete version of the non-linear $\sigma$-model
on  $S_F^2$ and were related to cyclic cohomology \cite{connes}. 

This article is organized as follows. In section 2, we will show how
projection operators correspond to solitons and show that in the limit
of large $j$, the soliton is really a smeared version of a point
particle on an ordinary sphere. In section 3, the geometry of the
moduli space of $k$ well-separated solitons is explored. The moduli
space is shown to be non-singular even when the solitons coincide. The
gradient term in the action lifts most of the moduli, and the
remaining moduli form an ordinary $S^2$. This gradient term provides
an attractive potential between the solitons, but vanishes in the
limit of large $j$. In section 4, the two soliton case is worked out
in detail and the interaction potential is calculated. In section 5,
we study the limit in which the fuzzy sphere is flattened in to the
noncommutative plane, and argue that all our results conform with the
known results for solitons on the noncommutative plane, modulo some
effects related to operator-ordering ambiguities. This suggests to us
that $S_F^2$ is a natural candidate for the ultraviolet regulated
version of the noncommutative plane. Our results are summarized in
section 6.

There is now a substantial body of work on $S^2_F$, starting from the
works of\cite{madore,gkp12}. Solitons and monopoles in non-linear
$\sigma$-models on $S_F^2$ were studied by \cite{bbvy} (see also
\cite{gkp3}), while topological issues such as instantons,
$\theta$-term and derivation of the chiral anomaly were discussed in
\cite{balvai}.  (For an alternate derivation of the chiral anomaly,
see \cite{presnajder}.) The continuum limit of the fuzzy non-linear
$\sigma$-model has been discussed in \cite{bamaoc}. The phenomenon of
UV-IR mixing for scalar theories on $S_F^2$ was first shown in
\cite{vaidya}, and further studied in \cite{doocpr,chmast}. Interest
in $S^2_F$ has increased since Myers showed that $D0$-branes in a
constant Ramond-Ramond field arrange themselves in the form of a fuzzy
sphere \cite{myers}. There have been investigations by \cite{alresc}
regarding open string versions of WZW models which naturally lead to
$S_F^2$. Gauge theories on $S_F^2$ have been studied by
\cite{gauges2}, while their continuum limits have been discussed by
\cite{iktw}. Noncommutative solitons on the fuzzy $S^2$ have appeared
in \cite{hinota} in the context of tachyon condensation and string
field theory. Further studies of topological as well as other issues
for fuzzy $S^2$ may be found in \cite{others2}.  Other aspects of
noncommutative solitons of the type discussed in \cite{gomist},
including connections to string theory, may be found in
\cite{othernc}.

\section{Solitons from Projectors}

The action for a single scalar field on fuzzy $S^2$ is given by
\begin{equation} 
S=\frac{1}{2j+1}{\rm Tr}_{{\cal H}^{(j)}}\Big(\dot{\Phi}^2 - [J_a, \Phi]^2 -
m^2 V[\Phi]\Big).
\label{scalaraction}
\end{equation} 
The field $\Phi$ is an arbitrary $(2j+1)\times (2j+1)$ hermitian
matrix, and the $J_a$ are the generators of the $SU(2)$ algebra in the
$(2j+1)$ dimensional representation, and ${\rm Tr}_{{\cal H}^{(j)}}$
is the trace is over the $(2j+1)$-dimensional Hilbert space ${\cal
H}^{(j)}$. We will call the term ${\rm Tr}_{{\cal H}^{(j)}}[J_a,
\Phi]^2$ as the {\it gradient} term, since in the continuum limit ($j
\rightarrow \infty$) it goes over to $\int ({\cal L}_a \Phi)^2$, where
${\cal L}_a = i \epsilon_{abc} x_a \partial_b$ are the vector fields
generating rotations on the 2-sphere.

In the limit $m^2 \rightarrow \infty$, the potential term of
(\ref{scalaraction}) gives the dominant contribution to the action
\cite{gomist,sprvol}. If the potential $V$ is polynomial, the it is
minimized by $\Phi=\lambda {\cal P}^{(k)}$ where the $\lambda$ is a
minimum of $V(x)$ and ${\cal P}^{(k)}$ is a hermitian projector of
rank $k$:
\begin{equation} 
{{\cal P}^{(k)}}^2 = {\cal P}^{(k)} = {{\cal P}^{(k)}}^{\dagger}.
\end{equation} 

Since this is a finite-dimensional matrix model, the rank $k$ of
non-trivial a projector satisfies $0<k<2j+1$. In an appropriate choice
of basis, a projector is simply a diagonal matrix with $k$ entries
being 1, the others being 0. The set of all rank $k$ projectors is
simply the Grassmannian $Gr(k,2j+1)$. 

When $m^2$ is large but finite, the gradient term of the action
(\ref{scalaraction}) must be taken into account as well. This term is
the energy of the configuration and provides a potential on the space
$Gr(k,2j+1)$. This potential lifts most of the moduli, but we will
argue that an $S^2$ worth of moduli remain as the lowest energy
configurations. The time dependent term ${\rm Tr}_{{\cal
H}^{(j)}}\dot{\Phi}^2$ of the action provides the dynamics, and thus
gives the metric on the lifted moduli space.

For simplicity, let us start with $k=1$. The set of all rank 1
projectors is the space ${\mathbb C}P^{2j}$, hence a rank 1 projector
is characterized by $4j$ moduli. The gradient term provides a
non-trivial potential on the moduli space ${\mathbb C}P^{2j}$. As a
result, not all configurations are equivalent: some configurations
have lower energy than others. The most general rank 1 projector is of
the form ${\cal P}^{(1)} = |Z\rangle \langle Z|$ where
\begin{equation} 
|Z \rangle = \sum_{\mu =-j}^{j} z_{\mu} |\mu \rangle, \quad z_{\mu}
 \in {\mathbb C} \quad {\rm and} \quad \langle Z|Z \rangle = 1,
\end{equation} 
the $\{|\mu\rangle \}$ being the standard angular momentum basis of
the $(2j+1)$-dimensional Hilbert space ${\cal H}^{(j)}$. 
  
We reproduce here the argument of \cite{sprvol} to find the set of
lowest energy configurations corresponding to the rank 1
projector. Rewriting the energy of the soliton as
\begin{equation} 
\frac{\lambda^2}{2j+1} {\rm Tr}_{{\cal H}^{(j)}} {\cal
P}^{(1)}[J_a,[J_a,{\cal P}^{(1)}]] =
\frac{\lambda^2}{2j+1}\Big(\langle Z|J_a J_a |Z\rangle - \langle Z|
J_a |Z \rangle \langle Z| J_a |Z\rangle \Big) = \frac{\lambda^2}{2j+1}
\langle Z|(\Delta \vec{J})^2 |Z\rangle,
\end{equation} 
we see that $|Z\rangle$ minimizes the energy if and only if it
minimizes the dispersion of $\Delta \vec{J}$, forcing it to be either
$|j\rangle$ and $|-j\rangle$, and thus value of energy to be $2j
\lambda^2 /(2j+1)$.

The most general (rank 1) projector of this energy is obtained by
applying rotations to, say, the state $|-j\rangle$ and using this to
construct the projector. This is simply $|\zeta\rangle\langle
\zeta|/\langle\zeta|\zeta\rangle$ where $|\zeta \rangle = e^{\zeta
J_+} |-j \rangle$ is the (non-normalized) $SU(2)$ coherent state. The
coordinate $\zeta$ has a simple interpretation: it is location of the
center-of-mass of the soliton, as we show below.

Corresponding to any operator ${\cal O}$ acting on the ${\cal
H}^{(j)}$, one can associate a function ${\cal O}(z,\bar{z})=\langle
z|{\cal O}|z \rangle / {\langle z|z\rangle}$ on the sphere, where $z$
is the stereographic coordinate. This is called the {\it covariant
symbol} \cite{berezin} of the operator ${\cal O}$. The covariant
symbol of the projector ${\cal P}_{(\zeta)}^{(1)} = |\zeta\rangle
\langle \zeta|/\langle \zeta|\zeta \rangle$ is the function
\begin{equation} 
{\cal P}^{(1)}_{(\zeta)}(z,\bar{z})=\frac{\langle z|\zeta\rangle \langle \zeta|
z \rangle}{\langle z|z \rangle}
\end{equation} 
For $\zeta=0$, the function
\begin{equation} 
{\cal P}^{(1)}_{(0)}(z, \bar{z}) = \frac{1}{(1+|z|^2/R^2)^{2j}} =
\cos(\theta/2)^{4j}, \quad {\rm where} \quad z = R
\tan(\theta/2)e^{i\phi}. 
\end{equation} 
may be interpreted as the soliton being at the north pole. For $j$
large, this function is strongly peaked around $\theta=0$, with a
spread of $(0, 2 {\tan}^{-1}(1/\sqrt{2j}))$, and almost zero outside
this region. It is in fact a regularized version of the
$\delta$-function on the sphere. We interpret this as a soliton of
angular size ${2\tan}^{-1}(1/\sqrt{2j})$ and located at
$\theta=0$. The dynamics of the single soliton is thus that of a
smeared point particle on the sphere.

Since it is only angular sizes that matter, we will restrict to $R=1$
henceforth whenever the fuzzy sphere is under consideration.

\section{Geometry of the moduli space}

Vectors in the $(2j+1)$-dimensional Hilbert space ${\cal H}^{(j)}$ are
usually expanded in terms of the basis $\{ |-~j\rangle, |-j+1\rangle,
\cdots |j\rangle \}$. One can also use a basis of $SU(2)$ coherent
states $\{|\zeta_1\rangle, |\zeta_2\rangle, \cdots,
|\zeta_{2j+1}\rangle \}$, where all the $\zeta_i$ are distinct points
on the the sphere. The coherent states we use are non-normalized:
$|\zeta_i\rangle = e^{\zeta_i J_+}|-j\rangle$. We show here that there
exists a non-singular basis even when some of the $\zeta_i$ are not
distinct.

Let us expand $|\zeta_i\rangle$ as 
\begin{equation} 
|\zeta_i\rangle = e^{\zeta_i J_+}|-j\rangle =
 \sum_{\mu=-j}^{j}\left[\frac{(2j)!}{(j+\mu)!(j-\mu)!}\right]^{1/2}
 \zeta^{j+\mu} |\mu\rangle \equiv \sum_{\mu=-j}^{j}c_\mu |\mu \rangle.
\end{equation} 
The basis $\{|\zeta_1\rangle, |\zeta_2\rangle, \cdots,
|\zeta_{2j+1}\rangle \}$ can be expressed in terms of the standard
basis $\{ |-j\rangle, |-j+1\rangle, \cdots |j\rangle \}$ as
\begin{equation}
\left(\begin{array}{c}|\zeta_1\rangle \\ 
		 |\zeta_2\rangle \\
		\vdots\\
	        |\zeta_{2j+1}\rangle
\end{array}\right) = 
\left(\begin{array}{ccccc}
c_{-j} &c_{-j+1}\zeta_1 &c_{-j+2}\zeta_1^2 &\cdots &c_j\zeta_1^{2j}\\
c_{-j} &c_{-j+1}\zeta_2 &c_{-j+2}\zeta_2^2 &\cdots &c_j\zeta_2^{2j}\\
\vdots & \vdots & \vdots & \vdots & \vdots \\
c_{-j} & c_{-j+1}\zeta_{2j+1} & c_{-j+2}\zeta_{2j+1}^2 &\cdots &
c_j\zeta_{2j+1} ^{2j} \\
\end{array}\right)
\left(\begin{array}{c}|-j\rangle \\
		|-j+1\rangle\\
		\vdots\\
		|j\rangle
\end{array}\right).
\end{equation}
The transformation matrix $U$ can be written as $V.C$ where V is the
Vandermonte matrix
\begin{equation} 
V=\left(\begin{array}{ccccc}
1 & \zeta_1 & \zeta_1^2 & \cdots & \zeta_1^{2j}\\
1 & \zeta_2 & \zeta_2^2 & \cdots & \zeta_2^{2j}\\
\vdots & \vdots & \vdots & \vdots & \vdots \\
1 & \zeta_{2j+1} & \zeta_{2j+1}^2 & \cdots & \zeta_{2j+1} ^{2j} \\
\end{array}\right) \quad {\rm and} \quad C = {\rm
diag}(c_{-j},\cdots,c_{j}). 
\end{equation} 
What happens when say, $\zeta_1 \rightarrow \zeta_2$? One can choose
$|\zeta_1\rangle$ and $(|\zeta_1\rangle - |\zeta_2 \rangle)/(\zeta_1
-\zeta_2)$ as basis vectors instead of $|\zeta_1\rangle$ and
$|\zeta_2\rangle$. The new basis has a well-defined limit even when
$(\zeta_1 -\zeta_2) \rightarrow 0$. In fact, it is easy to see that
$(|\zeta_1\rangle - |\zeta_2 \rangle)/(\zeta_1 -\zeta_2)$ tends to
$J_+|\zeta_1\rangle$. The vectors $|\zeta_1\rangle$ and
$J_+|\zeta_1\rangle$ are linearly independent, and so the basis $\{
|\zeta_1 \rangle, J_+ |\zeta_1 \rangle, |\zeta_3 \rangle, \cdots
|\zeta_{2j+1} \rangle \}$ is non-degenerate.

It is easy to see what happens when $\zeta_1,\cdots,\zeta_m$ coalesce
at the point $\zeta$. We choose as basis elements the vectors
$|\zeta\rangle, J_+\zeta\rangle, \cdots, J_+^m|\zeta\rangle$. These
are linearly independent, and along with the remaining distinct
$|\zeta_i\rangle$'s form a non-degenerate basis for our vector space.

The rank $k$ projector corresponding to the $k$-soliton configuration
is
\begin{equation} 
{\cal P}^{(k)}=\sum_{i,j=1}^k |\psi_i\rangle (h^{-1})_{ij}\langle\psi_j|,
\label{genkproj}
\end{equation}
where the $|\psi_i\rangle$ are linearly independent vectors in the
Hilbert space ${\cal H}$ and the matrix $h$ has entries $h_{ij}
=\langle\psi_i|\psi_j\rangle$. This projector projects onto the
subspace spanned by the vectors $|\psi_1\rangle, \cdots,
|\psi_k\rangle$. For describing well-separated solitons, we can write
this projector in terms of the coherent state basis as
\begin{equation} 
{\cal P}_{(\zeta_1, \cdots,\zeta_k)}^{(k)}=\sum_{i,j=1}^k
|\zeta_i\rangle (h^{-1})_{ij}\langle\zeta_j|
\label{kprojector}
\end{equation}
where $\zeta_1,\cdots, \zeta_k$ are $k$ points on the two-sphere. Any
permutation of these points gives us the same projector, and hence
this projector corresponds to a point on the space ${\cal M}_k =
\overbrace{{\mathbb C}P^1 \times \cdots {\mathbb C}P^1}^{k\;\;{\rm
copies}}/{\cal S}_k$, where ${\cal S}_k$ is the permutation group of
$k$ objects. Now, it well-known ${\cal M}_k$ is simply ${\mathbb
C}P^k$ (for a physicist's proof, see \cite{majbac}), so the
projector (\ref{kprojector}) is labeled by a point in ${\mathbb
C}P^k$. This is a K\"ahler manifold with the K\"ahler potential
$K(\zeta_a, \bar{\zeta}_b)$ given by
\begin{equation} 
K(\zeta_a,\bar{\zeta}_b) = \ln {\rm det}\;h
\label{kpot}
\end{equation} 
The metric $g_{\alpha \bar{\beta}}$ on the moduli space ${\cal M}_k$
is calculated from $K(\zeta_a,\bar{\zeta}_b)$ as 
\begin{equation} 
g_{\alpha \bar{\beta}} = \partial_\alpha
\partial_{\bar{\beta}}K(\zeta_a,\bar{\zeta}_b) 
\label{modulimetric}
\end{equation} 
Let us look at the rank 1 projector ${\cal
P}_{(\zeta)}^{(1)}=|\zeta\rangle \langle\zeta|/
\langle\zeta|\zeta\rangle$. The metric $g_{\alpha \bar{\beta}}$ on the
reduced moduli space comes from the kinetic energy term
\begin{equation}
\frac{\lambda^2}{2j+1}{\rm Tr}_{{\cal H}^{(j)}}\dot{{\cal
P}}_{(\zeta)}^{(1)}{}^2 = \lambda^2 g_{\alpha \bar{\beta}}
\dot{\zeta}_{\alpha}\dot{\bar{\zeta}}_\beta = \lambda^2
\left(\frac{2j}{2j+1}\right) \frac{2 |\dot{\zeta}|^2}{(1+|\zeta|^2)^2}.
\end{equation}
Thus $g_{\alpha \bar{\beta}}$ is simply the round metric on the
sphere, and the motion of the soliton can be thought of as the motion
of a free particle of mass $2j \lambda^2 /(2j+1)$ on the sphere. 

The gradient term in the the action (\ref{scalaraction}) gives the
energy of the $k$-soliton configuration. For $j$ finite, it leads
to attraction between the solitons and makes them clump together on
top of each other, as we will explicitly demonstrate for the case of
two solitons in the next section. It will be argued here that the
attraction becomes extremely weak in the limit of large $j$, and in
fact vanishes as $j\rightarrow \infty$.

Let us write the gradient term as
\begin{eqnarray} 
E[\zeta_a, \bar{\zeta_b}] &=& \frac{\lambda^2}{2j+1} {\rm Tr}_{{\cal
H}^{(j)}}{\cal P}_{(\zeta_1,\cdots,\zeta_k)}^{(k)}[J_a,[J_a, {\cal
P}_{(\zeta_1,\cdots,\zeta_k)}^{(k)}]] \\ \nonumber
&=& \lambda^2 \frac{2}{2j+1} {\rm
Tr}_{{\cal H}^{(j)}}\left(j(j+1){\cal P}_{(\zeta_1,\cdots,\zeta_k)}^{(k)} -
J_a {\cal P}_{(\zeta_1,\cdots,\zeta_k)}^{(k)} J_a {\cal
P}_{(\zeta_1,\cdots,\zeta_k)}^{(k)}\right) 
\end{eqnarray} 
Using the identities 
\begin{eqnarray} 
\langle \zeta_a|J_+|\zeta_b \rangle &=& \partial_{\zeta_b} h_{ab}, \quad
\langle \zeta_a|J_-|\zeta_b \rangle = \partial_{\bar{\zeta}_a} h_{ab},\\ 
J_3|\zeta_a\rangle &=& (-j + \zeta_a J_+)|\zeta_a \rangle ,
\end{eqnarray}   
the energy $E[\zeta_a, \bar{\zeta_b}]$ can be re-written as 
\begin{equation} 
E[\zeta_a, \bar{\zeta_b}] = \lambda^2 \frac{2}{2j+1} \left[ kj -
2j^2\sum_{a,b,c,d=1}^k [h_{ab} (h^{-1})_{bc} h_{cd} (h^{-1})_{da}]
\frac{(\bar{\zeta}_d \zeta_a + \bar{\zeta}_b \zeta_c - 2 \bar{\zeta}_d
\zeta_c)}{(1+\bar{\zeta}_d \zeta_a)(1 + \bar{\zeta}_b \zeta_c)}
\right]
\label{fullpotential}
\end{equation} 
It is not difficult to see that the product $[h_{ab} (h^{-1})_{bc}
h_{cd} (h^{-1})_{da}]$ goes to zero exponentially as $j \rightarrow
\infty$, so $E[\zeta_a, \bar{\zeta_b}] \rightarrow 2jk/(2j+1)$. In
other words, the energy of the multi-soliton configuration is a
constant independent of the locations of the solitons in the limit $j
\rightarrow \infty$.

For studying the bound state of $k$ solitons (where $k<j$), one uses
the basis $\{ |\zeta \rangle, J_+ |\zeta \rangle, \cdots, J_+^k
|\zeta\rangle \}$, and calculates the K\"ahler potential
(\ref{kpot}). With a little work, it can be shown that
\begin{equation} 
K(\zeta_a,\bar{\zeta}_b) = \ln (f(k,j)) + k(2j-(k-1))
\ln(1+|\zeta|^2),
\end{equation} 
where $f(k,j)$ is some function that is not relevant for the purpose
of calculating the metric. We can calculate $g_{\alpha \bar{\beta}}$
from the K\"ahler potential and see that the kinetic energy term
\begin{equation} 
\frac{\lambda^2}{2j+1} {\rm Tr}_{{\cal H}^{(j)}}\dot{{\cal
P}}^{(k)}{}^2 = \lambda^2 \frac{k(2j-(k-1))}{2j+1} \frac{2
|\dot{\zeta}|^2}{(1+|\zeta|^2)^2}
\end{equation} 
allows the interpretation of the bound state of $k$ solitons as a free
particle of mass $\lambda^2 k(2j-(k-1))/(2j+1)$ moving on the
$S^2$. For $k$ fixed and $j \rightarrow \infty$, the binding energy
$\lambda^2 (k-1)/(2j+1)$ vanishes, and the mass of the bound state is
the same as the mass of $k$ solitons of rank 1.

When $m^2 \neq \infty$, in general there are corrections to
(\ref{fullpotential}) coming from the potential term of
(\ref{scalaraction}) as well, which we have ignored in this
discussion. We briefly comment on these leaving the details for future
work. When $m^2$ is finite, fluctuations of the scalar field $\Phi$
need not be restricted to constant rank projectors. One expects such
fluctuations to change the number of solitons and hence play an
important role in the quantum field theory. Furthermore, interplay
between the limits $j \rightarrow \infty, R \rightarrow \infty$ and
$m^2 \rightarrow \infty$ hints at an interesting phase structure
deserving future exploration.

\section{Two-soliton case}

Let us look at the case of the rank 2 projector in detail. The
projector is of the form 
\begin{equation} 
{\cal P}_{(\zeta_1,\zeta_2)}^{(2)} = \sum_{i,j=1}^2 |\zeta_i\rangle
(h^{-1})_{ij}\langle\zeta_j|
\end{equation}
Corresponding to this projector is the function $\langle z|{\cal
P}^{(2)}|z \rangle /\langle z|z \rangle$, its covariant symbol. It is
given by
\begin{eqnarray} 
{\cal P}^{(2)}_{(\zeta_1, \zeta_2)}(z, \bar{z}) &=& \frac{1}{({\rm
det}\;\;h)(1+|z|^2)^{2j}} \Big[(1+\bar{z}\zeta_1)^{2j}
(1+\bar{\zeta}_1 z)^{2j} (1+|\zeta_2|^2)^{2j} \\ \nonumber
&&- (1+\bar{z}\zeta_1)^{2j}(1+\bar{\zeta}_2 z)^{2j} (1+\bar{\zeta}_1
\zeta_2)^{2j} - (1+\bar{z}\zeta_2)^{2j}(1+\bar{\zeta}_1 z)^{2j}
(1+\zeta_1 \bar{\zeta}_2)^{2j} \\ \nonumber 
&&+ (1+\bar{z}\zeta_2)^{2j} (1+\bar{\zeta}_2 z)^{2j}
(1+|\zeta_1|^2)^{2j}\Big]  
\end{eqnarray}   
where det $h = (1+|\zeta_1|^2)^{2j} (1+\zeta_2|^2)^{2j} - (1+\zeta_1
\bar{\zeta}_2)^{2j} (1+\bar{\zeta}_1 \zeta_2)^{2j}$. A 2-dimensional
plot of this function gives us an idea of how the two soliton
configuration looks like, which is plotted in the following
figure. One of the solitons is placed at the north pole ($\zeta=0)$,
while the other has spherical polar coordinates $(\theta,0)$.

\begin{figure}
\centerline{\epsfig{figure=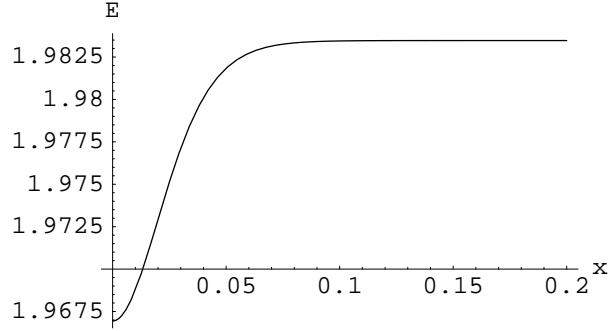,clip=2cm,width=8cm}}
\caption{Two-soliton energy $E(x)$ for j=60, where $x=|\zeta|^2$.}
\label{fig:2body}
\end{figure}

\begin{tabular}{ll}
\epsfig{figure=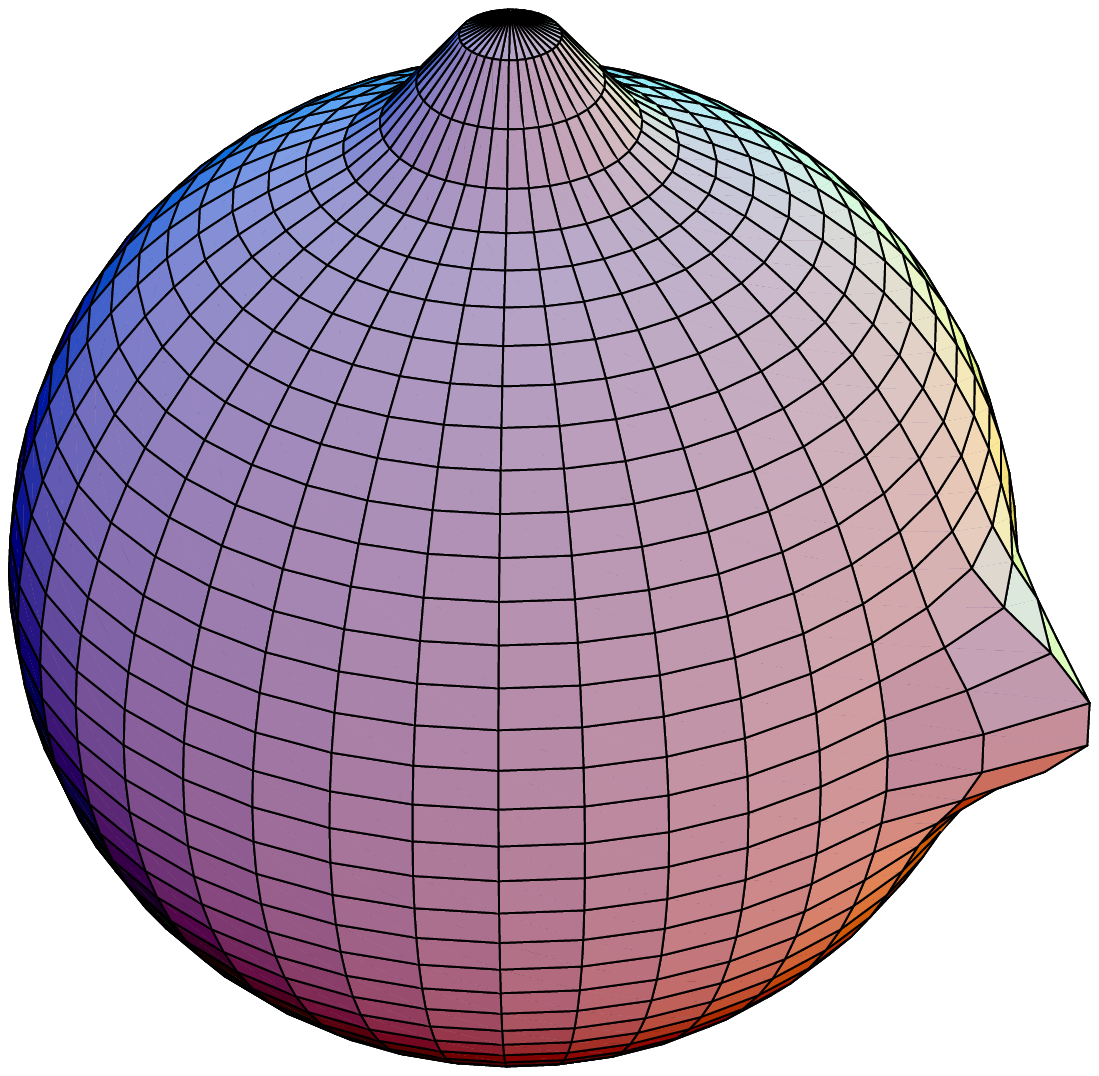,width=.4\textwidth} &
\epsfig{figure=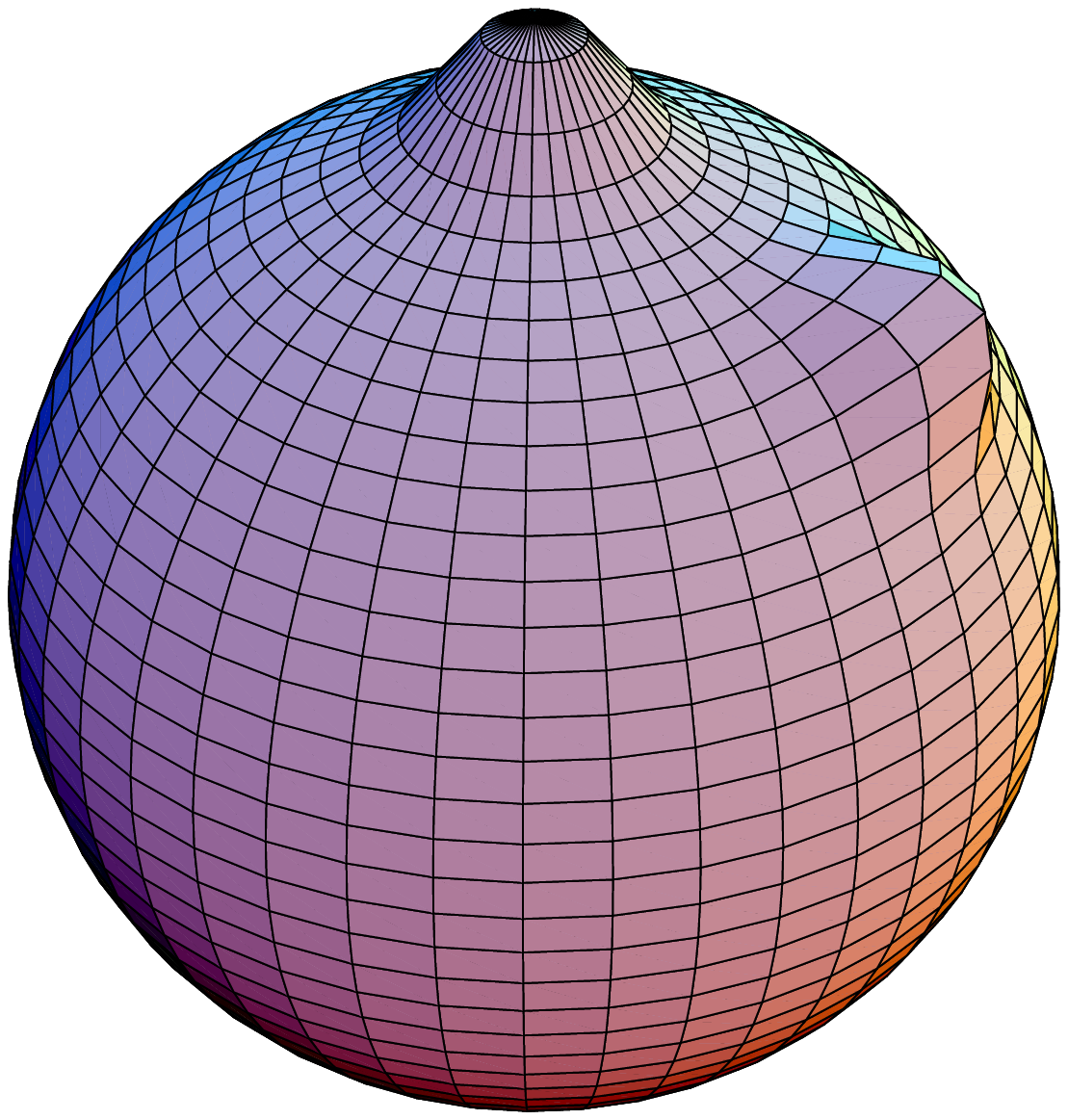,width=.4\textwidth} \\
(a) Angular separation $\theta =\pi/2.$ & (b) Angular separation
$\theta=\pi/4$. \\
\epsfig{figure=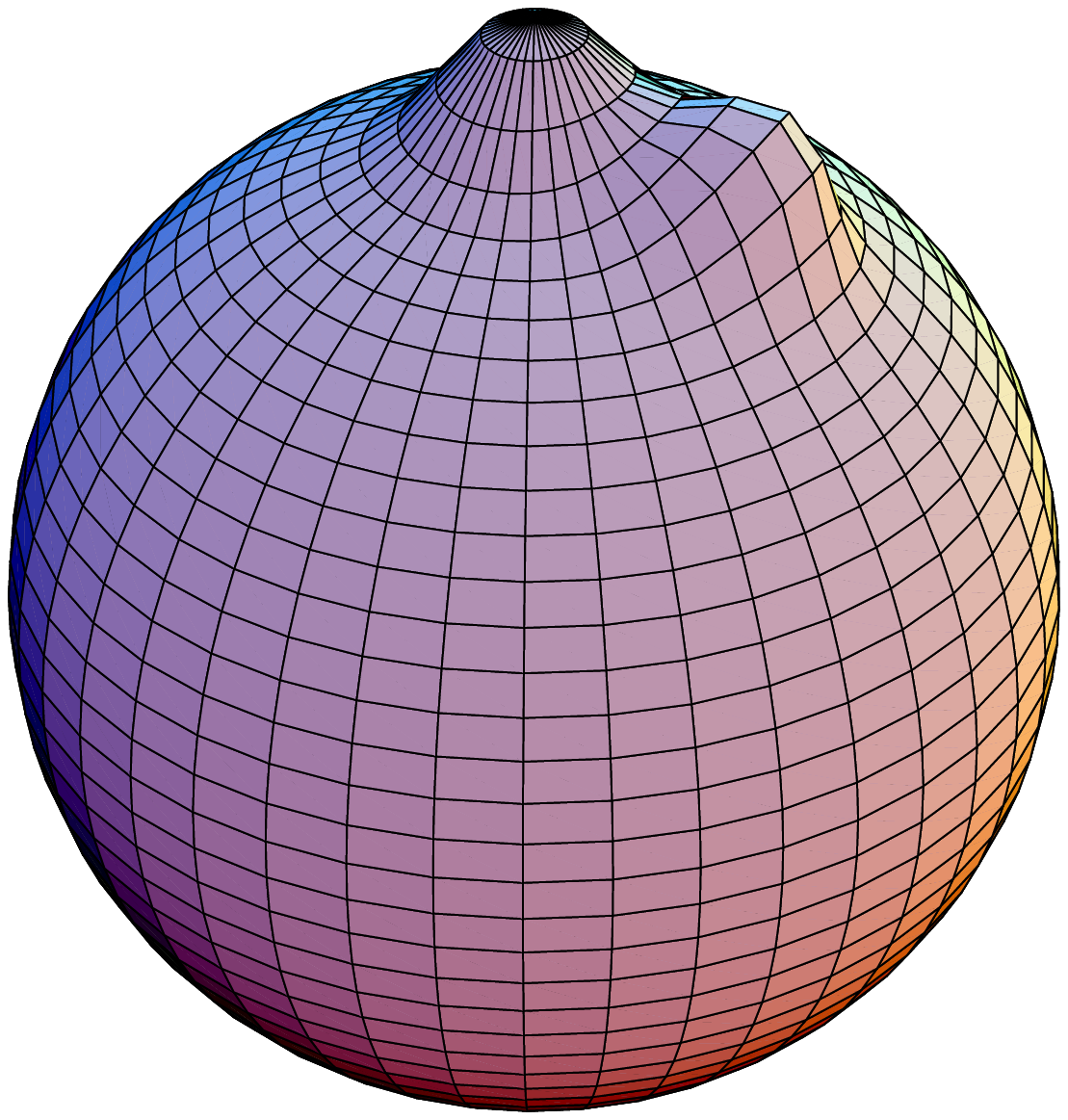,width=.4\textwidth} &
\epsfig{figure=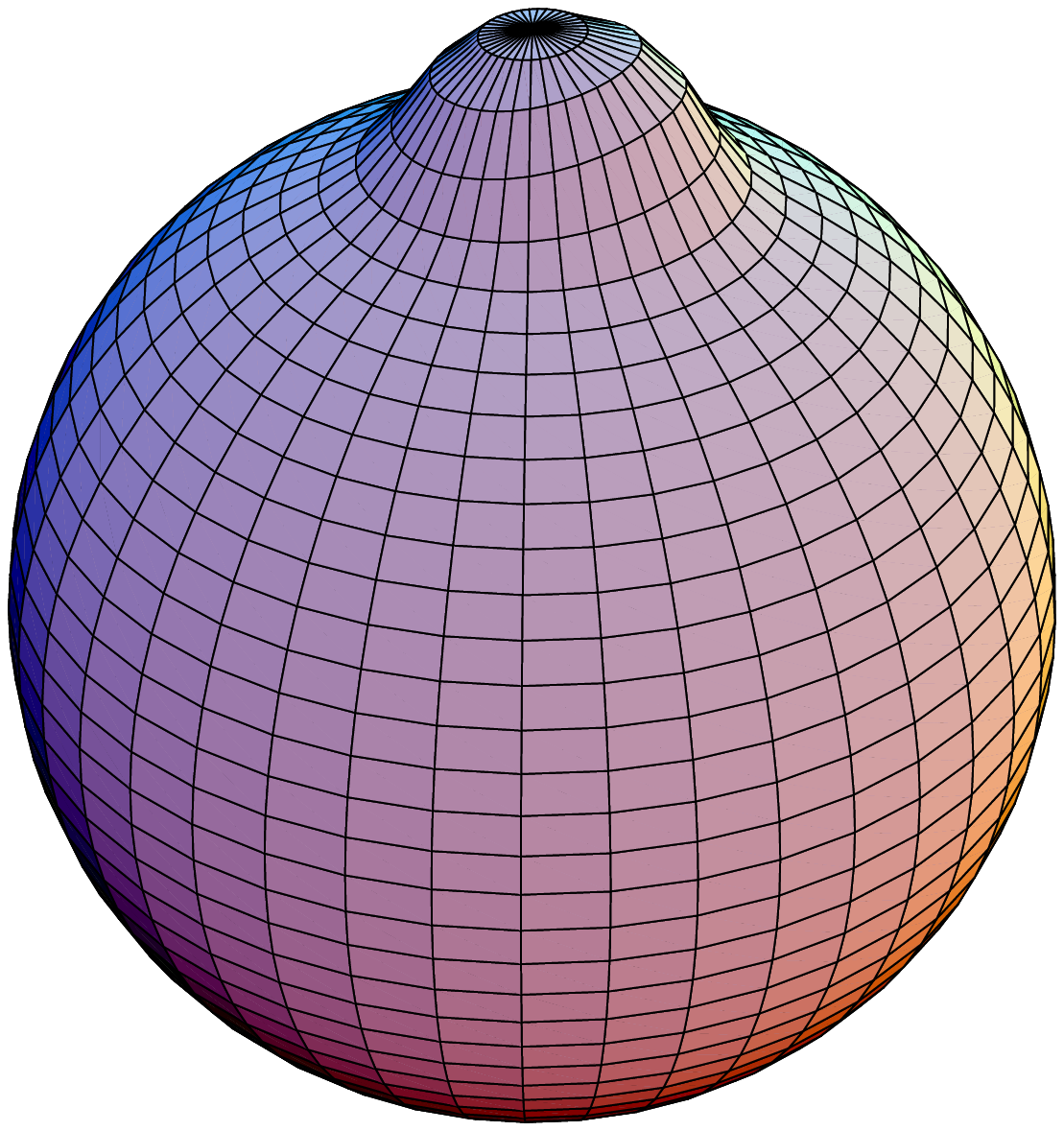,width=.4\textwidth}\\
(c) Angular separation $\theta=\pi/8$. & (d) Angular separation
$\theta=\pi/32$.\\ 
\multicolumn{2}{c}{Two solitons on a sphere of radius 5, with $j=60$.}
\end{tabular}
\vfil\eject To study the energy $E(\zeta_a, \bar{\zeta}_b)$ of the
configuration with solitons located at $\zeta_1$ and $\zeta_2$, it
should be remembered that since ${\cal P}^{(2)}_{(\zeta_1,\zeta_2)}$
is completely symmetric under $\zeta_1 \leftrightarrow \zeta_2$, a
good local coordinate for studying the coincident limit is $x=(\zeta_1
- \zeta_2)^2$. Without loss of generality, one of the solitons can be
put at 0 and the other at $\zeta$ to find $E$ as a function of $x$:
\begin{equation} 
E(x)=\lambda^2 \frac{4j}{2j+1}\left(1 -
\frac{2j(x^2(1+x)^{2j})}{(1+x)[(1+x)^{2j} - 1]^2}\right).
\end{equation} 
This is bounded, and has a smooth limits as $x \rightarrow 0$ or
$\infty$:
\begin{equation} 
E(0)=\lambda^2 \frac{2(2j-1)}{2j+1}, \quad E(\infty)=\lambda^2
\frac{4j}{2j+1}.
\end{equation}  
The function $E(x)$ is interesting because the $x \rightarrow 0$
and $j \rightarrow \infty$ limits do not commute. As long as $j$ is
finite, $E(0)$ is the global minimum. The potential is approximately
constant everywhere else. As $j \rightarrow \infty$, $E(0)$ approaches
the asymptotic value $4j\lambda^2/(2j+1)$ (We have plotted the
behavior of the function $E(x)$ in Fig.\ref{fig:2body} for $\lambda=1$).
Hence although the force between the solitons is attractive, it is
extremely weak for large values of $j$. The solitons move about freely
on the sphere, oblivious of each other's existence, unless they happen
to pass very close to each other, in which case they attract to form a
weak bound state. In fact, the binding energy between the solitons
vanishes as $j \rightarrow \infty$.

As observed earlier, the moduli space of two solitons is $Gr(2,2j+1)$
which is reduced to ${\mathbb C}P^2$ because of the gradient term of
the action. Using (\ref{modulimetric}), one can calculate the metric
$g_{\alpha \bar{\beta}}$ on this moduli space. The coordinates
$\zeta_1,\zeta_2$ are unconventional for describing ${\mathbb C}P^2$,
but are the natural choice in this context. The explicit form of the
metric is not very illuminating, but has a simple form for $j
\rightarrow \infty$:
\begin{eqnarray} 
g_{\zeta_1 \bar{\zeta_1}} &=& \frac{1}{(1+|\zeta_1|^2)^2}, \\
g_{\zeta_2 \bar{\zeta_2}} &=& \frac{1}{(1+|\zeta_2|^2)^2}, \\
g_{\zeta_1 \bar{\zeta_2}} &=& g_{\zeta_2 \bar{\zeta_1}} = 0.
\end{eqnarray} 
These limits are derived for $\zeta_1 \neq \zeta_2$. To understand the
limit of coincident solitons, we make a change of coordinates $y_1 =
(\zeta_1 + \zeta_2), y_2 = (\zeta_1 - \zeta_2)^2$, and expand in the
neighborhood of $y_2 = 0$. To order $|y_2|^2$, to find that:
\begin{eqnarray} 
g_{y_1 \bar{y_1}} &=& \left(\frac{2j-1}{2j+1}\right)
\frac{8}{(4+|y_1|^2)^2}, \\ 
g_{y_2 \bar{y_2}} &=& \left(\frac{2j-1}{2j+1}\right)\frac{8}{3}
\frac{(3|y_1|^2 + 8j -2)}{3(4+|y_1|^2)^2}, \\ 
g_{y_1 \bar{y_2}} &=& g_{y_2 \bar{y_1}} = 0. 
\end{eqnarray} 

The $g_{y_1 \bar{y_1}}$ component of the metric simply tells us that
the center-of-mass coordinate $y_1$ describes a sphere. The $g_{y_2
\bar{y_2}}$ component is non-vanishing, proving that the moduli space
is indeed smooth at the point $y_2 = 0$. 

The gradient term of (\ref{scalaraction}) lifts the modulus $y_2$, and
the bound state of two solitons is characterized by a single
coordinate $\zeta$, and the corresponding projector ${\cal
P}^{(2)}_{(\zeta)}$ can be read off from (\ref{genkproj}) with $|\psi_1
\rangle = |\zeta \rangle, |\psi_2 \rangle = J_+ |\zeta \rangle$. For
this projector, one finds that the metric on the moduli space is
\begin{equation} 
\frac{\lambda^2}{2j+1}{\rm Tr}_{{\cal H}^{(j)}}\dot{{\cal
P}}^{(2)}_{(\zeta)}{}^2 = \lambda^2 \frac{2(2j-1)}{2j+1}\frac{2
|\dot{\zeta}|^2}{(1+|\zeta|^2)^2}.  
\end{equation} 
This is a free particle of mass $2(2j-1)\lambda^2/(2j+1)$ on the
sphere.  The behavior of the many body interaction potential
(\ref{fullpotential}) is more intricate, but follows certain general
features. First of all, the interaction between the solitons is weak
for large $j$, and vanishes in the limit $j \rightarrow \infty$. For
finite $j$, in addition to the global minimum corresponding to all the
solitons being on top of each other, there are other extrema
corresponding to the solitons clumping at two antipodal points on the
sphere. More precisely, if there are $k$ solitons, then $k_1$ of these
form a bound state at, say, the north pole, while the remaining
$k-k_1$ form a bound state at the south pole. The global minimum of
the potential function corresponds to all $k$ solitons sitting on top
of each other to form a single bound state. The energy of this
configuration is $k(2j-(k-1))\lambda^2/(2j+1)$. When $k_1$ are at the
north pole and $k-k_1$ are at the south pole, the energy of the
configuration is $k_1(2j-(k_1-1))\lambda^2/(2j+1) +
(k-k_1)(2j-(k-k_1-1))\lambda^2/(2j+1)$. We also conjecture that there
are extrema when the solitons are placed at the corners of regular
solids that can be inscribed by a sphere.

\section{The flattening limit}
In the coherent state picture, it is very easy to flatten the fuzzy
sphere into the noncommutative plane. Following \cite{perelomov}, we
simply perform the replacements
\begin{equation} 
J_+ = \sqrt{2j} a^\dagger, \quad J_- = \sqrt{2j} a, \quad \zeta =
(2j)^{-1/2} \alpha,
\end{equation}
and take $j \rightarrow \infty$. Here $a,a^\dagger$ are the
annihilation and creation operators of the harmonic oscillator
algebra. Then, for example,
\begin{equation} 
|\zeta \rangle = e^{J_+ \zeta} |-j \rangle \rightarrow |\alpha \rangle
= e^{\alpha a^\dagger}|0\rangle 
\end{equation} 
Various quantities of interest can now be calculated in this limit. In
particular, we get
\begin{eqnarray} 
g_{\alpha_1 \bar{\alpha_1}} &=& \frac{1 - e^{-|\alpha_1 - \alpha_2|^2}
(1+|\alpha_1 - \alpha_2|^2)}{(1 - e^{-|\alpha_1 - \alpha_2|^2})^2}, \\
g_{\alpha_1 \bar{\alpha_2}} &=& \frac{e^{-2|\alpha_1 - \alpha_2|^2}
(1 - e^{|\alpha_1 - \alpha_2|^2}(1 - |\alpha_1 -
\alpha_2|^2)}{(e^{-|\alpha_1 - \alpha_2|^2} - 1)^2}, \\ 
g_{\alpha_2 \bar{\alpha_2}} &=& \frac{(1 - e^{-|\alpha_1 -
\alpha_2|^2}(1+|\alpha_1 - \alpha_2|^2))}{(1 - e^{-|\alpha_1 -
\alpha_2|^2})^2} 
\end{eqnarray}
Again, it is straightforward to see that the singularity at $\alpha_1 =
\alpha_2$ is a fake one, and that the various components of the metric
$g$ have a smooth limit as $\alpha_1 \rightarrow \alpha_2$. 

The $|\alpha \rangle$ are the (non-normalized) standard coherent
states of the Heisenberg-Weyl algebra that are used to study
noncommutative solitons on the plane. It is thus obvious that in this
limit, the results on $S_F^2$ go over to the results on the
noncommutative ${\mathbb R}^2$ as discussed in \cite{gohesp}.

A small puzzle arises when one studies the two-soliton configuration
in the limit of the noncommutative plane. In this limit, the
corresponding function is 
\begin{equation} 
{\cal P}^{(2)}_{(\alpha_1,\alpha_2)}(z, \bar{z}) = \frac{e^{-|z -
\alpha_1|^2} + e^{-|z - \alpha_2|^2}}{1 - e^{-|\alpha_1 -
\alpha_2|^2}} - \frac{e^{-(z-\alpha_1)(\bar{z} - \bar{\alpha}_2)} +
e^{-(z - \alpha_2)(\bar{z} - \bar{\alpha_1})}}{e^{|\alpha_1 -
\alpha_2|^2} - 1}.
\end{equation} 
In particular for $\alpha_1, \alpha_2 \rightarrow 0$, ${\cal
P}^{(2)}_{(0,0)}(z, \bar{z})$ has the form
\begin{equation} 
{\cal P}^{(2)}_{(0,0)}(z, \bar{z})=(1+|z|^2)e^{-|z|^2}
\end{equation} 
which is different (for small $z$) from the one obtained by
\cite{gohesp}. The resolution is not difficult: the Moyal-Weyl
transformation used in \cite{gohesp} corresponds to Weyl ordering,
whereas the covariant symbol corresponds to normal ordering. It is
well-known that functions corresponding to different ordering of
operators match at large distances (which is also true in this case),
but can differ for small distances (see for eg \cite{perelomov}).
Hence we see that while operator ordering issues are not important
while working at finite $j$, they certainly become relevant in the
noncommutative plane limit.

\section{Outlook}
Scalar theories on fuzzy $S^2$ admit finite energy configurations
constructed from rank $k$ projectors, which are localized lumps
(i.e. solitons) of size $\tan^{-1}(1/\sqrt{2j})$. The low energy
dynamics of these solitons is described by motion on the Grassmannian
$Gr(k,2j+1)$. Because the gradient term of the action provides a
non-trivial potential on $Gr(k,2j+1)$, this moduli space is
reduced. The solitons attract each other in general and the lowest
energy configuration for finite $j$ corresponds to all the solitons
being on top of each other. However, the attraction between the
solitons vanishes in the limit of large $j$, as does the binding
energy, suggesting that the solitons are like BPS particles. The
reduced moduli space corresponding to the $k$-soliton configuration is
${\mathbb C}P^k$ and is smooth: the apparent singularity corresponding
to the coalescence of several solitons is smoothed out by a different
choice of basis in the Hilbert space ${\cal H}^{(j)}$.

The limit corresponding to the noncommutative plane reproduces the
known results for the solitons on the noncommutative plane (modulo
considerations related to various kinds of operator ordering). With
hindsight, it should not be surprising that we have reproduced the
results for the noncommutative plane starting from the fuzzy
sphere. After all, ${\mathbb C}P^1$ looks locally like the complex
plane ${\mathbb C}$. The moduli space of $k$ solitons on the
noncommutative plane is ${\mathbb C}^{\otimes k}/{\cal S}_k \simeq
{\mathbb C}^k$ whereas that for solitons on the fuzzy sphere is
${\mathbb C}P^k$, which looks locally like ${\mathbb C}^k$. It thus
seems that the noncommutative sphere is an excellent candidate for the
ultraviolet (and simultaneously infrared!) regularization of the
noncommutative plane.

The construction of multi-soliton configurations on other fuzzy
manifolds (like the ones discussed in \cite{sprvol,abiy,bdlmo})
remains an open question. In particular, it would be interesting to
see if some of the features that we have discovered continue to hold;
in particular, whether the force between solitons vanishes in the
continuum limit.

{\bf Acknowledgments:} It is a pleasure to thank Nitin Nitsure for
discussions concerning algebraic geometry. This work was supported in
part by Department of Energy grant DE-FG03-91ER40674.

\bibliographystyle{unsrt}

\bibliography{paper}

\end{document}